\documentclass[twocolumn]{autart}    

\makeatletter
\newif\if@restonecol
\makeatother

\usepackage{graphicx}
\usepackage{amsmath,amsfonts,amssymb}
\usepackage{xcolor}
\usepackage[ruled,vlined,norelsize]{algorithm2e}

\makeatletter
\newenvironment{DCprotocol}[1][htb]{%
    \SetAlgorithmName{CFOR-REG Protocol}{}{List of Protocols}%
    \begin{algorithm}[#1]
    \SetAlgoLined
    \SetAlgoVlined
    \DontPrintSemicolon
     
}{%
    \end{algorithm}%
}
\makeatother

\makeatletter
\newenvironment{DUprotocol}[1][htb]{%
    \SetAlgorithmName{UFA-REG Protocol}{}{List of Protocols}%
    \begin{algorithm}[#1]
    \SetAlgoLined
    \SetAlgoVlined
    \DontPrintSemicolon
     
}{%
    \end{algorithm}%
}
\makeatother

\definecolor{myred}{rgb}{0.8, 0.0, 0.0}

\newcommand{\rea}{\mathbb{R}}
\newcommand{\nat}{\mathbb{N}}

\newcommand{\smax}[0]{\textsc{max}}
\newcommand{\smin}[0]{\textsc{min}}
\newcommand{\savg}[0]{\textsc{avg}}
\newcommand{\sint}[0]{\textsc{int}}

\DeclareMathOperator*{\pr}{Prob}

\usepackage{xcolor}

\usepackage{enumitem}

\newcommand{\G}{\mathcal{G}}
\newcommand{\V}{\mathcal{V}}
\newcommand{\E}{\mathcal{E}}
\newcommand{\M}{\mathcal{M}}
\newcommand{\X}{\mathcal{X}}

\newcommand{\virg}[1]{``#1"}
\newcommand{\norm}[1]{{\left\vert\kern-0.3ex\left\vert #1\right\vert\kern-0.3ex\right\vert}}
\newcommand{\mnorm}[1]{{\left\vert\kern-0.3ex\left\vert\kern-0.3ex\left\vert#1\right\vert\kern-0.3ex\right\vert\kern-0.3ex\right\vert}}
\newcommand{\abs}[1]{{\left\vert #1 \right\vert}}

\makeatletter
\let\c@author\relax
\makeatother
\usepackage[citestyle=numeric-comp,
	 bibstyle=ieee,
	 giveninits=true,
      maxbibnames=99,
	 ]{biblatex}

\AtEveryBibitem{
	\clearfield{issn}
	\clearfield{isbn}
	\clearfield{archivePrefix}
	\clearfield{arxivId}
	\clearfield{pmid}
	\clearfield{doi}
	\clearfield{eprint}
	\clearfield{note}
	\clearfield{publisher}
    \clearfield{organization}
	\ifentrytype{online}{}{\ifentrytype{misc}{}{\clearfield{url}}}
	\ifentrytype{book}{\clearfield{doi}}{}
}

\begin{document}

\begin{frontmatter}


\title{Algebraic Connectivity Control and Maintenance in Multi-Agent Networks under Attack\thanksref{footnoteinfo}} 

\thanks[footnoteinfo]{This work was supported in part by the project Research on Interdisciplinary Issues in Mechatronics Integration (YJSJ24001) funded by the Fundamental Research Funds for the Central Universities and the Innovation Fund of Xidian University, and in part by the MUR National Recovery and Resilience Plan funded by the European Union - NextGenerationEU under project SERICS (PE00000014).}

\author[author1,author2]{Wenjie Zhao}\ead{wnjzhao@stu.xidian.edu.cn},
\author[author2]{Diego Deplano}\ead{diego.deplano@unica.it},             
\author[author1,author3]{Zhiwu Li}\ead{zhwli@xidian.edu.cn},  
\author[author2]{Alessandro Giua}\ead{giua@unica.it},
\author[author2]{Mauro Franceschelli}\ead{mauro.franceschelli@unica.it}  

\address[author1]{SEME, Xidian University, Xi'an, China}          
\address[author2]{DIEE, University of Cagliari, Cagliari, Italy}
\address[author3]{ISE, Macau University of Science and Technology, Taipa, Macau}


\begin{abstract}
%
This paper studies the problem of increasing the connectivity of an ad-hoc peer-to-peer network subject to cyber-attacks targeting the agents in the network.
The adopted strategy involves the design of local interaction rules for the agents to locally modify the graph topology by adding and removing links with neighbors. 
Two distributed protocols are presented to boost the algebraic connectivity of the network graph beyond $k-2\sqrt{k-1}$ where $k\in\nat$ is a free design parameter; these two protocols are achieved through the distributed construction of random (approximate) regular graphs.
One protocol leverages coordinated actions between pairs of neighboring agents and is mathematically proven to converge to the desired graph topology. The other protocol relies solely on the uncoordinated actions of individual agents and it is validated by a spectral analysis through Monte-Carlo simulations.
Numerical simulations offer a comparative analysis with other state-of-the-art algorithms, showing the ability of both proposed protocols to maintain high levels of connectivity despite attacks carried out with full knowledge of the network structure, and highlighting their superior performance.
\end{abstract}

\end{frontmatter}

\section{Introduction}
\label{sec:introduction}

An ad-hoc multi-agent network refers to an ensemble of agents communicating with each other directly through wireless links, without the need for a central infrastructure or relying on any pre-existing infrastructure, characterized by its variability over time.
This kind of network arises naturally in different applications, including the maintenance of sensor networks~\cite{sharma2020distributed,meena2023efficient}, query access optimization in massive data networks~\cite{36}, and multi-robot coordination~\cite{37}, among others.
In such applications, it is crucial to address perturbations such as sudden disconnections of agents due to failures~\cite{b7,b8,b9} or attacks carried out by malicious agents~\cite{b10}. 
One of the most critical scenarios to avoid is the disconnection of the network into disjoint components, which can severely disrupt the information flow throughout the network. 
To quantify the connectivity of a graph, various metrics have been introduced, primarily focusing on the minimum number of nodes and edges that must be removed to render the graph disconnected. 
Established metrics for this purpose include algebraic connectivity and the Fiedler eigenvector~\cite{b11,b12}, the Kirchhoff index~\cite{b13,b14}, and the edge/node expansion ratio~\cite{b15}. 
These metrics effectively reflect the connectivity quality of a graph and are instrumental in evaluating its robustness and synchronizability~\cite{b17}.

Various algorithms according to different connectivity measures designed~\cite{b18,b19,b20,b21} to improve graph connectivity have recently gained significant attention. 
Adding edges to a graph is a naive approach to increase connectivity but is not practical in many applications where each edge represents a virtual or physical link between the corresponding agents. 
For example, when the edges represent physical communication channels~\cite{b22}, a large number of edges is often not desirable due to increased costs.

Random regular graphs represent an interesting class of graphs that score high on several connectivity measures while maintaining a low number of edges~\cite{b16,b23}. A graph is said to be $k\text{-regular}$ if each node has a degree equal to $k$. A $k\text{-regular}$ graph is said to be random if it is selected uniformly at random from the set of all $k\text{-regular}$ graphs with the same number of nodes. 
Yaz{\i}c{\i}o{\u{g}}lu \emph{et al.} progressively developed a series of distributed protocols to enhance the robustness of network, transforming any connected graph into a connected random regular graph.
The study in~\cite{b24} employs decentralized degree regularization by minimizing node degree variances, thereby ensuring uniformity and stability.
Further advancing their method, a distributed reconfiguration scheme proposed in~\cite{b25} constructs random regular graphs via self-organization, maintaining connectivity and average degree while further randomizing edges.
Additionally, a decentralized method introduced in~\cite{Yazicioglu2015formation} that facilitates the transformation of any connected interaction graph with an initial average degree $d_{\savg} > 2$ into a connected random $k\text{-regular}$ graph for some $k \in [d_{\savg},d_{\savg}+2]$. 
This approach not only enhances the robustness of the network but also preserves a comparable number of edges to those in the original graph, thereby showcasing the inherent adaptability and resilience of these graph structures.
Building on previous works, the study in~\cite{Zohreh22} introduces a novel distributed protocol that enables a network of cooperative agents to self-organize into an inexact of random $k$-regular graph, where all but one node achieve a degree of $k$.

\textbf{The main contribution of this paper} is the design of two distributed protocols to continuously self-organize a network, by local modifications of the graph topology, to boost its connectivity. In particular, they allow controlling the algebraic connectivity of the corresponding Laplacian matrix, which can be increased, exceeding a lower bound of $k - 2 \sqrt{k-1}$ -- with $k\in\nat$ being a free design parameter -- by constructing random (approximate) $k$-regular graphs.
The two proposed protocols differ in the aspects of both methodology and performance.
One protocol leverages coordination among pairs of agents, which allows us to present a rigorous formal analysis to prove the convergence of the network topology toward an exact random $k$-regular graph.
The other protocol relies solely on uncoordinated actions taken locally by the agents -- thus benefitting from an easier communications scheme -- and presents improved performance in terms of actions to be taken to achieve the desired topology. 
This comes at the cost of resembling the communication graph into an approximated (and not exact) random $k$-regular graph, as shown via a spectral analysis of the Laplacian matrix with Monte-Carlo simulations.
Numerical simulations corroborate the theoretical and experimental findings, showing how both protocols are able to increase the algebraic connectivity and maintain it at high levels despite changes in the network size and topology due attacks causing node and link failures.

{\bf Structure of the paper.} Preliminaries are introduced in Section~\ref{sec:preliminaries}.
In Section~\ref{sec:protocols} two distributed protocols to control the algebraic connectivity of a graph through its rewiring are proposed.
Section~\ref{sec:numerical} presents numerical simulations to validate the proposed protocols and compare them with the state-of-the-art. 
Concluding remarks are given in Section~\ref{sec:conclu}.


\section{Preliminaries on graph theory and networks} \label{sec:preliminaries}

A multi-agent system (MAS) consists of multiple interacting agents whose pattern of interactions can be modeled by a \emph{graph} ${\G=(\V,\E)}$, where $\V=\left\{1,\ldots, n\right\}$ is the set of \emph{nodes}, representing the agents, and $\E\subseteq \V\times \V$ is the set of \emph{edges} connecting the nodes, representing the point-to-point communication channels between the agents.
Graphs are assumed to be undirected (i.e., if $(i,j)\in \E$ then $(j,i)\in \E$) and without self-loops (i.e., $(i,i) \not\in E$ for all $i\in\V)$.
A \emph{path} between two nodes $i,j\in \V$ is a sequence of consecutive edges ${(i,p),(p,q),\ldots,(r,s),(s,j)}$, where each edge shares a node with its predecessor.
An undirected graph $\mathcal{G}$ is said to be \emph{connected} if there exists a path between any pair of nodes $i,j \in \V$.
Nodes $i$ and $j$ are said to be \emph{neighbors} if there is an edge between them, i.e., $(i,j)\in \E$.
The set of neighbors of node $i$ is denoted by $\mathcal{N}_i=\left\{j\in \V: (i,j)\in \E\right\}$.
Similarly, the set of \emph{$2\text{-hops neighbors}$} is denoted by $\mathcal{N}_i^2$, which includes only agents $j$ such that there exists a path between $i$ and $j$ of exactly $2$ edges.
The \emph{degree} of a node $i$ is the number of neighbors, denoted by $d_i = \abs{\mathcal{N}_i}$, where $\abs{\cdot}$ denotes the cardinality of a set.
Consequently, minimum, maximum, and average degrees of the graph $\mathcal{G}$ are denoted by
$
d_{\smin}(\G)=\min_{i\in \V} d_i$, $d_{\smax}(\G)=\max_{i\in \V} d_i$, $d_{\savg}(\G) =   {\sum_{i\in \V}}\frac{d_i}{n}$.
The \emph{degree matrix} $D=\text{diag}(d_1,\ldots,d_n)$ is a matrix whose diagonal elements are the node degrees while all off-diagonal elements are zero.
The \emph{adjacency matrix} $A_n=\{a_{i,j}\}\in\{0,1\}^{n\times n}$ encodes the graph $\G$ and $a_{i,j}=1$ if and only if $(i,j)\in\E$, and $a_{i,j}=0$ otherwise.
The eigenvalues of the adjacency matrix are denoted by $\mu_i$ with $i=1,\ldots,n$. 
The \emph{Laplacian matrix} of graph $\G$ is defined as $L_n = D-A_n\in\nat^{n\times n}$ and its eigenvalues are denoted by $\lambda_i$ with $i=1,\ldots,n$ in~\cite{bullo2018lectures}.
We assume the eigenvalues are sorted in ascending order, i.e., $\mu_i\leq \mu_{i+1}$ and ${\lambda_i\leq \lambda_{i+1}}$ for $i=1,\ldots,n-1$.
The second smallest eigenvalue $\lambda_2$ of the Laplacian matrix $L$ is known as the \textit{algebraic connectivity} of the graph.



A specific class of graphs known to enhance robustness and resilience is that of \emph{random regular graphs}, which is introduced next. In the rest of this section, we detail how random regular graphs can be obtained via a random process associated with a Markov chain.

\begin{defn}\label{def:regraph}
${\G}$ is said to be \virg{$k\text{-regular}$} if all nodes have the same degree $k\in\nat$. Moreover,
a $k\text{-regular}$ connected graph is said to be \virg{random} if it is selected uniformly at random from all $k\text{-regular}$ connected graphs with the same number of nodes. 
\end{defn}

Note that a $k$-regular graph may not exist for arbitrary values of $n$ and $k$: in particular, the product $nk$ must be even (we make the standing assumption that $n\geq 2$ and $k\in[2,n-1]$). 
This motivates us to define approximate regular graphs, given any $n$ and $k$.

\begin{defn}
${\G}$ is said to be \virg{$\text{approximate $k$\text{-regular}}$} if all nodes have a degree within $\{k,k+1\}$.
\end{defn}

A random $k\text{-regular}$ graph is known to have the largest eigenvalue equal to $k$. All the 
other eigenvalues, as conjectured by Alon~\cite{b16} and proved by Friedman~\cite{b23}, are bounded by $2\sqrt{k-1}$ with high probability when the graph is picked at random among all $k\text{-regular}$ graphs, which allows the finding a lower bound on the algebraic connectivity. We summarize these results in the following proposition.
\begin{prop}\label{pro:specgap}
    Given a connected random $k\text{-regular}$ graph, the eigenvalues $\mu_1\leq \cdots \leq\mu_{n-1} \leq\mu_n$ of its adjacency matrix $A_n$ satisfies, with high probability (see~\cite[Theorem 1.1]{b23}),
    $
    \mu_n=k$, $\max\{\abs{\mu_{n-1}},\abs{\mu_1}\}\leq 2 \sqrt{k-1}.
    $
    Thus, the second smallest eigenvalue $\lambda_2$ of the Laplacian matrix $L_n$ (i.e., the algebraic connectivity of the corresponding graph), is lower-bounded with high probability 
    $$\lambda_2\geq \lambda_{2,lb} : = k-2\sqrt{k-1}.$$
\end{prop}
In this work we consider \emph{unstructured} networks, whose graph topology is not fixed but it may change over time. In particular, we allow the agents to perform local modifications by opening/closing communication channels with other agents. These events take place at discrete times indexed by ${t=0,1,2,\ldots}$, thus giving rise to a time varying graph $\G (t)=(\V,\E (t))$ with a fixed set of nodes and a time-varying set of edges.
We call this a \emph{random graph process} (RGP) which, starting with an initial graph $\G(0)$, generates a sequence of graphs $\{\G(0),\G(1),\G(2),\ldots\}$. By construction, the graph $\G(t+ 1)$ generated at time $t+1$ only depends on the graph $\G(t)$, but from a graph $\G(t)$ there may be multiple different graphs that can be generated, with different probabilities.
An RGP can generate only a finite set of graphs since the set of nodes is finite, denoted by $\mathbb{G}_n$.


One can associate to an RGP a finite discrete-time Markov chain $\M = (\X,P,\pi_0)$ such that a walk in $\X$ corresponds to a realization of the RGP, where $\X$ is the finite \emph{set of states} representing all the possible graphs with $n$ nodes, i.e., each state $X_i\in \X$ uniquely correspond to a graph in $\G_i\in \mathbb{G}_n$; $P$ is the \emph{transition probability matrix} representing the probability of transitioning from two different states, i.e., $p_{i,j}\in[0,1]$ represents the probability of generating graph $\G_j$ from $\G_i$; $\pi_0$ is the \emph{initial probability distribution}, i.e., $\pi_0(i)=1$ if $\G(0)=\G_i$, $X(0)=X_i$ and $\pi_0(i)=0$ otherwise. With this notation, a sequence of states $\{X(0),X(1),X(2),\ldots\}$ corresponds to a  sequence of graphs $\{\G(0),\G(1),\G(2),\ldots\}$.

A subset of $\X$ is called a \emph{component} if all states within it are \emph{recurrent}, i.e., from any node in the component is possible to reach all others.
A component $\X_{abs} \subseteq \X$ is defined as \emph{absorbing} if it is not possible to reach states outside the component starting from one within the component, i.e., given $j\notin \X_{abs}$, $p_{i,j}=0$ for all $i\in\X_{abs}$.
We now state an important property of Markov chains that have only one absorbing component.

\begin{prope}\label{pro:abscomp}
Let $(\X,P,\pi_0)$ be a Markov chain with only one absorbing component $\X_{abs}\subseteq \X$. A sequence of states  $X(t)$ almost surely hits $\X_{abs}$ in finite time and remains within it at subsequent times, i.e., 
$$\pr\left(\exists t^\star>0,\ \forall t\geq t^\star: X(t) \in \X_{abs}\right) = 1.$$
\end{prope}
An absorbing component is said to be \emph{ergodic} if it is \emph{aperiodic}, i.e., the maximum common divisor among the lengths of all cycles within it is one.
A Markov chain is \emph{ergodic} if it has a single  ergodic component.
%
%
\begin{prope}\label{pro:stadis}
Given an ergodic Markov chain $(\X,P,\pi_0)$, there exists a stationary probability distribution $\pi_s$ such that the probability of being on a state ${X_i\in\X}$ is equal to $\pi_s(i)$, i.e.,
$$\lim_{t\rightarrow \infty} \pr (X(t) = X_i) = \pi_s(i),\qquad \forall X_i\in \X.$$
\end{prope} 

We are now in the position to formally define the concept of \virg{uniform $k$-regularity} for RGPs, which clarifies what kinds of graphs are of interest in this paper, i.e., those that self-organize persistently into random $k$-regular connected graphs as in Definition \ref{def:regraph}.

\begin{defn}\label{def:kregularRGP}
An RGP whose initial graph $\G(0)$ is connected with $n$ nodes, is said to be \virg{uniformly $k$-regular} if the associated Markov chain $\M=(\X,P,\pi_0)$ is ergodic with ${\X^\star\subseteq \X}$, where $\X^\star$ represents the set of all $k$-regular connected graphs with $n$ nodes, i.e., each state $X_i\in \X$ uniquely corresponds to a connected regular graph, and the stationary probability distribution $\pi_s$ ensures that for all $X_i \in \X^\star$ it holds that $\pi_s(i) = |\X^\star|^{-1}.$
\end{defn}
This means that, a \virg{uniformly $k$-regular} RGP generates connected random $k$-regular graphs as $t\rightarrow\infty$ according to Definition \ref{def:regraph}.
%
%

\section{Proposed distributed protocols} \label{sec:protocols}
This section characterizes two different protocols to control the algebraic connectivity in unstructured networks by constructing regular graphs. 
In particular, we present:
\begin{itemize}
    \item the CFOR-REG Protocol in Section \ref{sec:algo_1}. It exploits coordination between pairs of neighboring agents, and is proven to steer the network topology toward a random regular graph;
    \item the UFA-REG Protocol in Section \ref{sec:algo_2}. It only relies on independent decision of the agents and it is proven to steer the network topology toward an approximate regular graphs, whose algebraic connectivity has been empirically proven to converge to that of random regular graphs for large networks.
\end{itemize}
Additionally, in Section \ref{sec:specdisttheo} we compare the spectral properties of the graph obtained with the two protocols with that of random regular graphs via Monte Carlo simulations.

\begingroup
\begin{DCprotocol}[!t]
\caption{\textbf{D}stributed \textbf{C}oordinated \textbf{F}ormation of \textbf{R}andom $k\text{-\textbf{R}egular}$ connected \textbf{G}raphs} \label{alg:regalgo_M}
\LinesNumbered
\SetKwInput{KwInput}{Input}
\SetKwInput{KwOutput}{Output}
\KwInput{A connected graph $\G=(\V,\E)$, $\epsilon, \beta \in (0, 1)$, and the desired integer degree $k\geq 2$}
\KwOutput{A random $k$-regular graph}
\For{$t = 1, 2, 3, \ldots$}{
    Each node, $i \in \V$, is actived with probability $1-\epsilon$\;
    \mbox{Each activated node $i \in \V_a$ picks a random $j \in \mathcal{N}_{i}$}\;
    \vspace{-13pt} 
    For each $i \in \V$, $R_{i} = \{i^{'} \in \mathcal{N}_{i} | i^{'} \text{ picked } i\}$\;
    {\ForEach{$(i, j)$ s.t. $i \in R_{j}$, $j \in R_{i}$, $d_{i} \geq d_{j}$}{
        $i$ picks at random a rule $r\in \{r_1, r_2, r_3, r_4\}$\;
        \uIf(\textcolor{gray}{\scriptsize \tcp*[f]{Rule 1}}){$r = r_{1}, |R_{i}| \geq 2, |R_{j}| \geq 2$}{
            $i$ picks at random a node $h \in R_{i} \setminus \{j\}$\;
            $j$ picks at random a node $f \in R_{j} \setminus \{i\}$\;
            \If{$(i, f) \notin \E, (j, h) \notin \E$}{
                \mbox{$\E \gets (\E \setminus \{(i, h), (j, f)\}) \cup \{(i, f), (j, h)\}$\;}
            }
        }
        \uElseIf(\textcolor{gray}{\scriptsize \tcp*[f]{Rule 2}}){$r = r_{2}, d_{i} > d_{j}, |R_{i}| \geq 2$}{
                $i$ picks at random a node $h \in R_i \setminus \{j\}$\;
                \If{$(j, h) \notin \E$}{
                    $i$ picks at random a value $\beta^{'} \in [0, 1]$\;
                    \uIf{$\beta^{'} > \beta$}{
                        $\E \gets (\E \setminus \{(i, h)\}) \cup \{(j, h)\}$\;
                    } \Else {
                        $\E \gets \E \cup \{(j, h)\}$\;
                    }
                }
            }
        \uElseIf(\textcolor{gray}{\scriptsize \tcp*[f]{Rule 3}}){$r = r_{3}, d_{i} > d_{j}, |R_{i}| \geq 2$}{
                $i$ picks at random a value $\beta^{'} \in [0, 1]$\;
                \If{$\beta^{'} > 1-\beta$}{
                    $i$ picks at random a node $h \in R_{i} \setminus \{j\}$\;
                    \If{$(j, h) \in \E$}{$\E \gets \E \setminus \{(i, h)\}$}
                }
            }
        \ElseIf(\textcolor{gray}{\scriptsize \tcp*[f]{Rule 4}}){$r = r_4$}{
                \uIf{$d_{i} > k, d_{j} > k, \mathcal{N}_j\cap \mathcal{N}_i\neq \emptyset$}{
                    $\E \gets \E \setminus \{(i, j)\}$\;
                } \ElseIf{$d_i < k$}{
                    $\mathcal{N}_{ji} :=  \{h \in \mathcal{N}_{j} \setminus (\mathcal{N}_{i} \cup \{i\})\}$\;
                    $i$ picks at random a node $h \in \mathcal{N}_{ji}$\;
                    \If{$d_{h} < k$}{
                        $\E \gets \E \cup \{(i, h)\}$\;
                    }
                }
            }  
        }
    }
}
\end{DCprotocol}
\endgroup

\subsection{Distributed formation of connected random $k$-regular graphs}\label{sec:algo_1}

The CFOR-REG Protocol is presented, which allows for the distributed formation of random $k$-regular graph exploiting coordination between neighboring agents, where $k$ is an arbitrary even natural number.

\begin{figure}[!t]
    \centering   \includegraphics[width=0.4\textwidth]{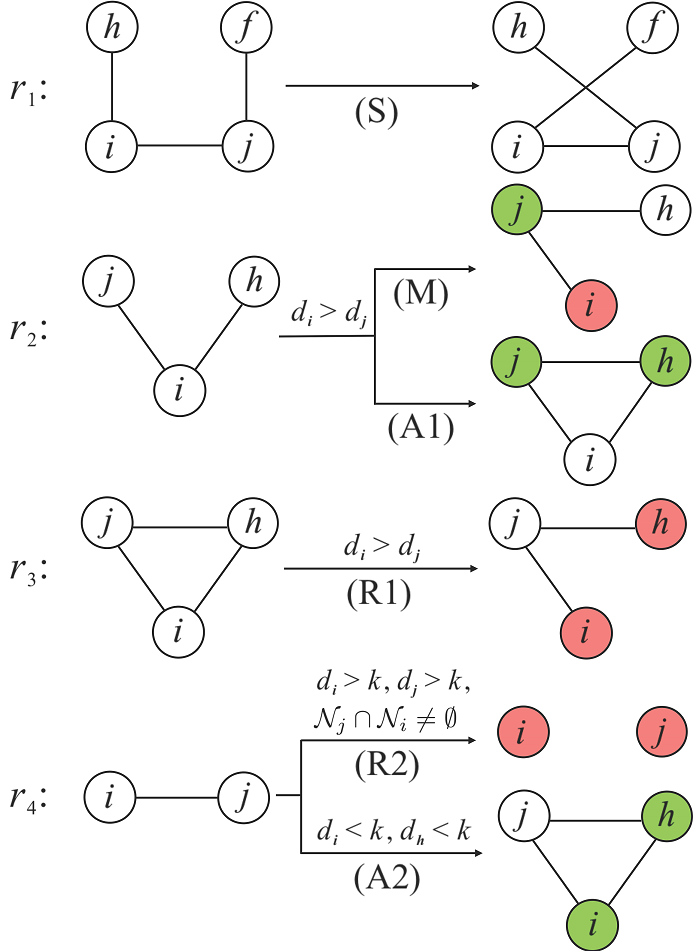}
    \caption{Rules and operations available in the CFOR-REG Protocol: nodes colored in red have their degree decreased, while nodes colored in green have their degree increased.}
    \label{fig:Algorithm_1}
\end{figure}

In accordance with the CFOR-REG Protocol, the nodes behave as follows: At each iteration $t=1,2,3,\ldots$, any node is activated with a probability $\epsilon\in(0,1)$ and picks one of its neighbors $j\in\mathcal{N}_i$ uniformly at random. Active nodes broadcast to their neighbors information about their degree and which neighbor they picked randomly; thus each node $i$ can build the list of nodes $R_i$ that has randomly picked itself, i.e.,
$
R_i = \{j\in \mathcal{N}_i: j \text{ has picked } i\}$, for all $i\in\V.
$
Each matched pair $(i,j)$ such that $i\in R_j$ and $j\in R_i$, ordered by $d_i\geq d_j$, has at its disposal the following set of operations:
\begin{itemize}
    \item \textbf{Swap (S)}: this operation randomizes the edges in the graph. Nodes $i,j$ swap two neighbors $h\in R_i\setminus(\mathcal{N}_j\cup\{j\})$ and $f\in R_j\setminus(\mathcal{N}_i\cup\{i\})$, i.e., edges $(i,h)$, $(j,f)$ are removed and edges $(i,f)$, $(j,h)$ are added.
    \item \textbf{Move (M)}: this operation balances the nodes' degree. If node $i$ has a degree greater than node $j$, then $i$ loses a neighbor $h\in R_i\setminus(\mathcal{N}_j\cup\{j\})$ that becomes of $j$, i.e., edge $(i,h)$ is removed and edge $(j,h)$ is added.
    \item \textbf{Add (A)}: this operation increases the number of edges in the graph, and may happen in two cases.
    \begin{itemize}
        \item \textbf{(A1)}: If node $i$ has a degree greater than node $j$, then a neighbor $h\in R_i\setminus(\mathcal{N}_j\cup\{j\})$ also becomes a neighbor of $j$, i.e., edge $(j,h)$ is added.
        \item \textbf{(A2)}: If a node $i$ has degree lower than $k$, then a neighbor $h\in\mathcal{N}_j\setminus\{i\}$ of the neighbor ${j\in\mathcal{N}_i}$ becomes a neighbor of $i$ if its degree $d_h$ remains lower than $k$, i.e., edge $(i,h)$ is added.
    \end{itemize}
    \item \textbf{Remove (R)}: this operation reduces the number of edges in the graph, which may happen in two cases.
    \begin{itemize}
        \item \textbf{(R1)}: If node $i$ has a degree greater than its neighbor $j$, then $i$ loses a common neighbor $h\in (R_i\cap\mathcal{N}_j)\setminus\{j\}$, i.e., edge $(i,h)$ is removed.
        \item \textbf{(R2)}: If nodes $i,j$ have degrees greater than $k$ and $\mathcal{N}_j\cap \mathcal{N}_i\neq \emptyset$, then $i$ and $j$ stop being neighbors, i.e., edge $(i,j)$ is removed. 
    \end{itemize}
\end{itemize}
These operations are executed by selecting at random one of the following rules, as illustrated in Fig.~\ref{fig:Algorithm_1}.
\begin{itemize}
    \item \textbf{Rule~1}: if $\abs{R_i}\geq 2$ and $\abs{R_j}\geq 2$, the pair tries to swap two neighbors by executing operation~(S);  
    \item \textbf{Rule~2}: if $\abs{R_i}\geq 2$, the pair either tries (with probability~$1-\beta$) to move an edge by executing operation~(M), or tries (with probability~$\beta$) to add an edge by executing operation~(A1);
    \item \textbf{Rule~3}: if $\abs{R_j}\geq 2$, the pair tries to remove an edge by executing operation~(R1);
    \item \textbf{Rule~4}: the pair either tries to add an edge by executing operation~(A2) or tries to remove an edge by executing operation~(R2).
\end{itemize}
%
%
\begin{thm}\label{thm:cfor-reg}
Consider an RGP due to the execution of the CFOR-REG Protocol starting from a connected graph $\G(0)$ with $n\geq k$ nodes and even $k\in\{4,6,8,\ldots\}$. Then:
\begin{itemize}
    \item[(i)] the graph $\G(t)$ remains connected at each time $t\geq 0$;
    \item[(ii)] there almost surely exists a $t^\star \in \nat$ such that the graph $\G(t)$ is $k\text{-regular}$ for all times $t\geq t^\star$, i.e.,
    $$ \pr(\exists t^\star>0, \forall t\geq t^\star : \G(t) \text{ is $k$-regular }) = 1;$$
    \item[(iii)] the RGP is \virg{uniformly $k$-regular}, i.e., it  generates connected random $k$-regular graphs as $t\rightarrow\infty$;
    \item[(iv)] the algebraic connectivity $\lambda_2(t)$ of the graph $\G(t)$ is, in expectation, greater than $k-2\sqrt{k-1}$ as $t\rightarrow \infty$.
\end{itemize}
\end{thm}
\begin{pf}
Statement $(i)$ is trivial, which can be directly verified in a graphical way by looking at Figure \ref{fig:Algorithm_1}, where all operations result in a connected subgraph. Then, the entire graph remains connected.

Statement $(ii)$ can be proved via Markov chain theory. Let $\M=(\X,P,\pi_0)$ be the Markov chain associated to the RGP, i.e.: $\X$ represents all the possible graphs with $n$ nodes; $P=\{p_{i,j}\}$ is the matrix whose elements represent the probability of transitioning between any two graphs; $\pi_0$ is the \emph{initial probability distribution}, i.e., $\pi_0(i)=1$ if $X_i\in\X$ is the state associated to graph $\G(0)$ and $\pi_0(i)=0$ otherwise. With this notation, statement $(ii)$ is equivalent to asking that the subset of states corresponding to $k$-regular connected graphs, denoted by $\X^{reg}$, is the only absorbing component of the Markov chain (according to Property \ref{pro:abscomp}).
To this aim, consider that any initial state $X(0)\notin\X^{reg}$ does not correspond to a $k$-regular connected graph, i.e., $2\abs{\E} \neq k \abs{\V}$; since $k$ is assumed to be even, for any number of nodes $\abs{\V}=n$, there always exists a number of edges satisfying $2\abs{\E}=k\abs{\V}$, a necessary condition for a $k$-regular graph. For the sake of simplicity, let us denote by $\G=\G(0)$ the initial graph and by $\G'=\G(t)$ a graph reached by a random walk on the Markov chain after $t$ transitions.
Then, either one of the following two cases may occur. 

$\textbf{Case 1}$: $2\abs{\E} < k \abs{\V}$, i.e., there are fewer edges than required, resulting in some nodes having degree less than $k$. We prove that $\G'$ can increase the number of edges by considering the only two possible scenarios:
    \begin{enumerate}
        \item $\G$ is $m$-regular with $m<k$: then $d_i=m$ for all $i\in \V$. Moreover, there surely exists a pair of nodes $(i,j)\in \E$ such that ${\mathcal{N}_{ij}=\mathcal{N}_i\setminus\{\mathcal{N}_j\cup\{j\}\}\neq \emptyset}$. Indeed, if $\mathcal{N}_{ij}=\emptyset$ for all pairs of nodes $(i,j)\in \E$, then the graph must be complete, i.e., $m=n$, which is in contradiction with the assumption $m<k$ thanks to $k< n$. Thus, an extra edge can be added by executing operation~(A2) in Rule~4.
        \item $\G$ is not $m$-regular with $m \leq k$: then there surely exists a pair of nodes $(i,j)\in\E$ such that $d_i>d_j$ and $\mathcal{N}_i\setminus\{\mathcal{N}_j\cup\{j\}\}\neq \emptyset$, i.e, $i$ must have at least one neighbor that is not linked to $j$. Thus, an extra edge can be added by executing operation~(A1) in Rule~2.
    \end{enumerate}
$\textbf{Case 2}$: $2\abs{\E} > k \abs{\V}$, i.e., there are more edges than needed, leading to some nodes having degree greater than $k$. We now prove that $\G'$ can decrease the number of edges. By~\cite[Lemma 4.8]{Yazicioglu2015formation} the execution of Rules~{1-2-3} can transform any graph $\G$ into a graph $\G''=(\V,\E'')$ containing at least one triangle, i.e., there is a triplet $(i,j,h)$ such that $\{(i,j)(j,h),(h,i)\}\subseteq \E''$. 
Thus, we consider $\G''$ as the new initial graph and analyze the only two possible scenarios:
\begin{enumerate}
        \item $\G''$ is $m$-regular with $m>k$: then $d_i=m$ for all $i\in \V$. Moreover, there necessarily exists a pair of nodes $(i,j)\in \E$ such that $\mathcal{N}_j\cap \mathcal{N}_i\neq \emptyset$ since there is at least one triangle. Thus, an edge can be removed by executing operation~(R2) in Rule~4.
        \item $\G''$ is not $m$-regular with $m \geq k$: then either all triangles have nodes with the same degrees, or there is at least one triangle with nodes of different degrees. If the latter case holds, then an edge can be removed by executing operation~(R1) in Rule~3. In the former case, we show that any triangle whose nodes have the same degree can be transformed into a triangle with nodes of different degrees. Consider a generic triangle composed of nodes $(i,j,h)$ such that $d_i=d_j=d_h\geq 2$.
        Consider the shortest path from any node of this triangle to a node whose degree is different from the degrees of the triangle nodes. Without loss of generality, let this path be between $i$ and $\ell$, and realize that all nodes in this path have the same degree that is equal to $d_i$, while $d_\ell \neq d_i$. Either one of the following cases occurs:
        \begin{itemize}
            \item[-] $d_\ell>d_i$: in this case, by executing Rule~2, node $\ell$ can always add an edge from one of its neighbors to the node on the shortest path, thus increasing its degree and, in turn, shortening the shortest path from $i$ to a higher degree node. By induction, the degree of node $i$ can be eventually increased, i.e., the triangle $(i,j,h)$ becomes such that $d_i>d_j=d_h$.
            \item[-] $d_\ell<d_i$: in this case, we first note that $d_i\geq 3$ since $i$ has two neighbors within the triangle and another neighbor on the shortest path. Therefore, the neighbor $p$ of $\ell$ on the shortest path has at least a neighbor $q$ outside the shortest path. Due to $d_p>d_\ell$ by assumption, by executing Rule~2 with operation~(M), it is always possible to remove the edge $(p,\ell)$ and add the edge $(q,\ell)$, thus decreasing the degree of node $p$ and, in turn, shortening the shortest path. By induction, the degree of node $i$ can be eventually decreased, i.e., the triangle $(i,j,h)$ becomes such that $d_i<d_j=d_h$.
        \end{itemize}
    \end{enumerate}
Hence, in either case, the edges of the network can be added or removed via the CFOR-REG Protocol until the number of edges implies an integer average degree equal to $k$.
Assume now that the graph is still not $k$-regular, i.e., the set $\V_{>k}$ of nodes with degree greater than $k$ and the set $\V_{<k}$ of nodes with degree greater lower $k$ are not empty. 
Consider two nodes $i\in\V_{>k}$ and $j\in\V_{<k}$ of minimum distance $\delta_{ij}\in\nat$ and the following cases:
\begin{enumerate}
    \item[(a)] $\delta_{ij}=1$, i.e., nodes $i$ and $j$ are neighbors. In this case, node $i$ has at least a neighbor $h$ that is not a neighbor of $j$ (as $i$ has a higher degree than $j$). Therefore, node $h$ can perform the Move (M) operation in Rule 2 to remove the edge with node $i$ and to add the edge to node $j$, thus yielding both sets $\V_{>k}$ and $\V_{<k}$ to lose one node.
    \item[(b)] $\delta_{ij}>1$, i.e., in the shortest path between $i$ and $j$ there are $d_{ij}-1$ nodes of degree exactly $k$ (since we are considering $i$ and $j$ of minimum distance). Let $p$ and $q$ be the first two nodes in the shortest path from $i$ to $j$. In this case, node $i$ has at least a neighbor $h$ that is not a neighbor of $j$ (as $i$ has higher degree than $j$). Thus, node $i$ and its closest neighbor $p$ on the shortest path to $j$ can execute the Swap (S) operation in Rule 1 and exchange neighbors $h$ and $q$, thus yielding a reduction of the distance $\delta_{ij}$ by $1$.
\end{enumerate}
Case (b) can be repeated until case (a) holds, which yield both sets $\V_{>k}$ and $\V_{<k}$ to lose one node. Thus, these sets eventually become empty, i.e., the graph is reshaped into a $k$-regular graph.
Once a $k$-regular graph is achieved, only the Swap (S) operation in Rule~1 can be performed, thus maintaining the degree of all nodes equal to $k$.
This proves that the subset of states $\X^{reg}$ corresponding to all the $k$-regular connected graphs is the only absorbing component, thus completing the proof of statement $(ii)$.

Statement $(iii)$, according to Definition \ref{def:kregularRGP}, corresponds to the case in which the Markov chain is ergodic on $\X^{reg}$, and the stationary probability distribution (which exists by Property \ref{pro:stadis}) is uniform. As proved in statement $(ii)$, once a random walk enters the absorbing component $\X^{reg}$ only the Swap (S) operation in Rule~1 can be executed.
This rule was originally presented in \cite{mahlmann2005peer} as the \virg{$1$-Flipper operation}, where it has been proven that, starting from any connected $k$-regular graph with $n$ nodes, in the limit for $t\rightarrow\infty$ this operation constructs all connected $k$-regular graphs with the same probability \cite[Theorem 1]{mahlmann2005peer}.

This can be verified by noticing that: 1) any $k$-regular connected graph can be constructed, i.e., all states in $\X^{reg}$ are recurrent; 2) there is a non-zero probability of remaining in the same graph, i.e., the transition probability matrix of the Markov chain has strictly positive diagonal entries and, in turn, all states in $\X^{reg}$ is aperiodic; 3) for any two graphs $\G$ and $\G'$ the probability of constructing $\G'$ from $\G$ is the same of constructing $\G$ from $\G'$, i.e., the transition probability matrix of the Markov chain is symmetric and, in turn, the stationary probability distribution is uniform.
From 1) and 2) it follows that the Markov chain achieves ergodicity on $\X^{reg}$ and thus it admits a stationary probability distribution $\pi_s$. From 3) it follows that $\pi_s(i) = |\X^{reg}|^{-1}$ for all graphs. Thus, according to Definition \ref{def:kregularRGP}, the RGP is uniformly $k$-regular, i.e.,
it generates connected random $k$-regular graphs as $t\rightarrow\infty$ according to Definition \ref{def:regraph}, completing the proof of statement $(iii)$.

Statement $(iv)$ holds since the RGP is uniformly $k$-regular and the algebraic connectivity of a random $k$-regular connected graph is lower bounded in expectation by $k-2\sqrt{k-1}$, see Property \ref{pro:specgap}. 
\hspace*{\fill}$\qed$
\end{pf}

\subsection{Distributed uncoordinated formation of approximate $k$-regular graphs}\label{sec:algo_2}
The UFA-REG Protocol enables the distributed formation of a random $k$-regular graph, and does not require coordination between neighboring agents.
Consequently, operations like the \virg{Swap} (S) of the CFOR-REG Protocol are not allowed now, since they would require coordination between pair of neighboring nodes. 
In order to allow the agent to locally modify the graph topology in a completely autonomous way while also avoiding conflicting actions, we introduce next the concept of \virg{\textit{ownership of a graph's edge}}.

We limit the freedom of nodes to remove and add edges by assigning to each edge $(i,j)\in\E$ a unique owner between nodes $i$ and $j$. Thereafter, each edge can be removed only by its owner, and the owner of each newly added edge $(i,j)$ is designated as the node responsible for its addition.
We model this scenario by means of a directed \emph{ownership graph} $\G_d=(\V,\E_d)$, which is a directed version of the undirected communication graph $\G=(\V,\E)$ defined as follows: for any edge $(i,j)\in \E$, either $(i,j)\in\E_d$ or $(j,i)\in\E_d$, denoting whether $i$ or $j$ owns the edge, respectively.
With this notation, the set of out-neighbors of node $i$, denoted by $\mathcal{N}_{i,out}=\left\{j\in \V: (i,j)\in \E_d\right\}$, represents the subset of neighbors $\mathcal{N}_i$ from which node $i$ can remove edges between them.

\begin{figure}[!h]
    \centering\includegraphics[width=0.4\textwidth]{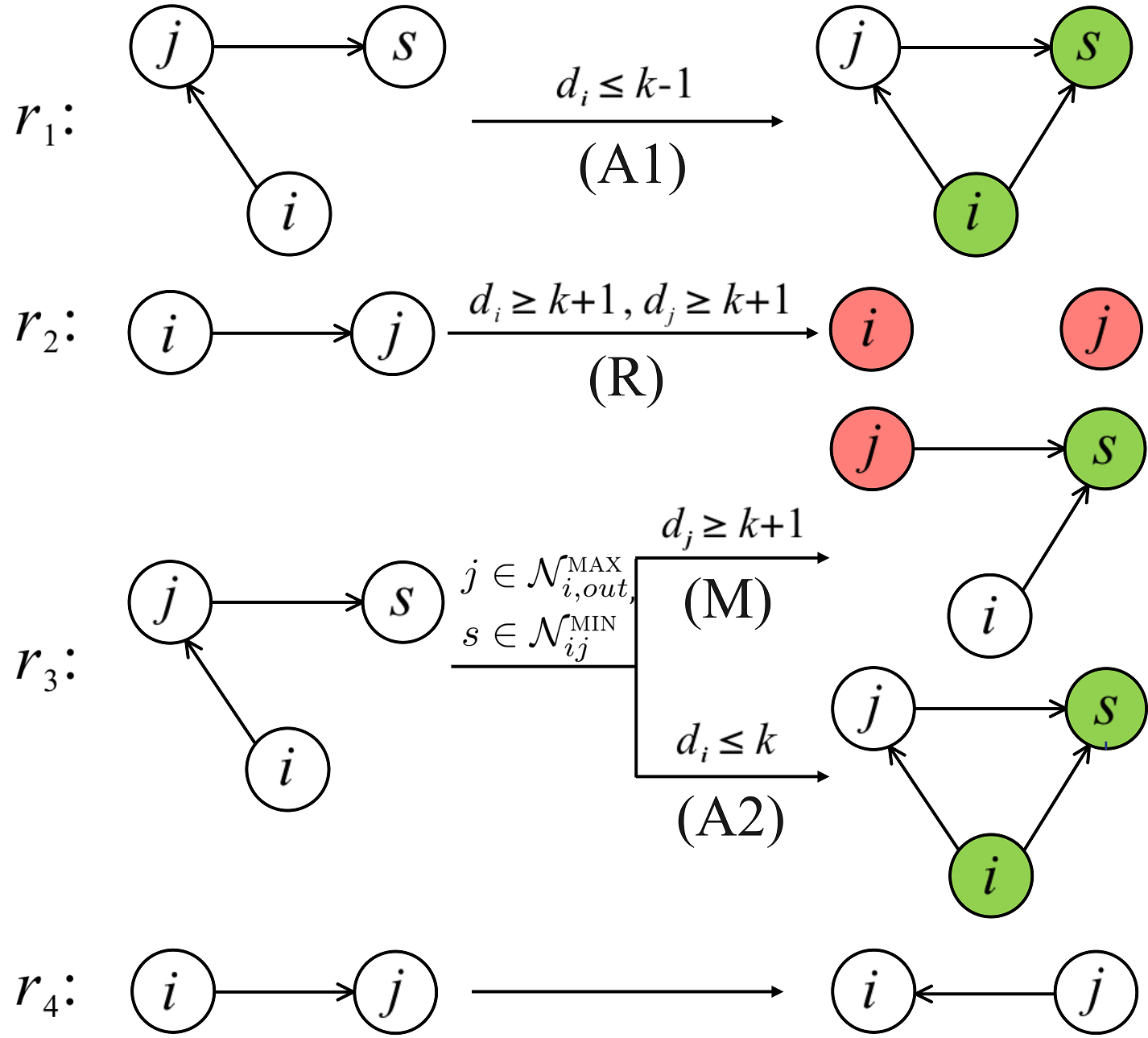}
    \caption{Rules and operations available in UFA-REG Protocol: nodes colored in red have their degree decreased, while those in green have their degree increased.}
    \label{fig:Algorithm_2}
\end{figure}

At each iteration each node is active with a probability $\epsilon \in (0,1)$ and locally modifies the network topology by executing one of the rules in $\Phi_{2}$ which are comprised by following operations:
\begin{itemize}
    \item \textbf{Move (M)}: this operation balances the degree of nodes and also at randomizing the edges in the graph. Node $i$ removes an edge from a neighbor $j\in\mathcal{N}_{i,out}$ with the highest degree such that $d_j\geq k +1$. Then, node $i$ adds an edge to a $2$-hop neighbor ${s\in \mathcal{N}_j\setminus\{\mathcal{N}_i\cup\{i\}\}}$ with lowest degree such that $d_s\leq \max\{k,d_j-1,d_i-1\}$.
    \item \textbf{Add (A)}: this operation increases the number of edges in the graph, and may happen in two cases.
    \begin{itemize}
        \item \textbf{(A1)}: If node $i$ has a degree $d_i\leq k-1$, then it adds an edge to a $2$-hop neighbor ${s\in \mathcal{N}_j\setminus\{\mathcal{N}_i\cup\{i\}\}}$, where $j\in \mathcal{N}_i$.
        \item \textbf{(A2)}: If node $i$ has a degree $d_i\leq k$, then it adds an edge to a $2$-hop neighbor ${s\in \mathcal{N}_j\setminus\{\mathcal{N}_i\cup\{i\}\}}$ with the lowest degree such that $d_s\leq k$, where $j\in\mathcal{N}_{i,out}$ has the highest degree such that $d_j\leq k$.
    \end{itemize}
    \item \textbf{Remove (R)}: this operation reduces the number of edges in the graph. If node $i$ has a degree $d_i\geq k+1$, then it removes an edge it owns from a neighbor $j\in \mathcal{N}_{i, out}$ with the highest degree such that $d_j\geq k+1$.
\end{itemize}

\begingroup
\begin{DUprotocol}[!t]
\caption{\textbf{D}stributed \textbf{U}ncoordinated \textbf{F}ormation of \textbf{A}pproximate $k\text{-\textbf{Re}gular}$ \textbf{G}raphs} \label{alg:regalgo}
\LinesNumbered
\SetKwInput{KwInput}{Input}
\SetKwInput{KwOutput}{Output}
\KwInput{A connected graph $\G=(\V,\E)$, $\epsilon, \beta \in (0, 1)$, and the desired integer degree $k\geq 2$}
\KwOutput{A random $\text{approximate}$ $k$-regular graph}
\For{$t = 1, 2, 3, \ldots$}{
        Each node, $i \in \V$, is activated with probability $1-\epsilon$\;
        Pick at random a node $i\in \V_a$\;
        \While(\textcolor{gray}{\scriptsize \tcp*[f]{Rule 1}}){$d_i \leq k-1$}{
            \For{$j \in \mathcal{N}_{i}$}{
                $\mathcal{N}_{ij} :=  \{s \in \mathcal{N}_{j} \setminus (\mathcal{N}_{i} \cup \{i\})\}$\;
                \If{$\mathcal{N}_{ij} \neq \emptyset$}{
                    $i$ picks at random a node $s \in \mathcal{N}_{ij}$\;
                    $\E \gets \E \cup \{(i,s)\}$\;
                    break \tcp*[h]{exit the for loop}\;
                }
            }
        }
        Node $i$ picks at random a rule $r\in \{r_2, r_3, r_4\}$\;
        \uIf(\textcolor{gray}{\scriptsize \tcp*[f]{Rule 2}}){$r=r_2$, $d_i \geq k + 1$}{
            \If{$\mathcal{N}_{i,out} \neq \emptyset$}{
                Node $i$ picks at random a node $j \in \mathcal{N}_{i,out}$\;
                \If{$d_j \geq k+1$}{
                    $\E \gets \E \setminus \{(i, j)\}$\;
                }
            }
        }
        \uElseIf(\textcolor{gray}{\scriptsize \tcp*[f]{Rule 3}}){$r=r_3$, $\mathcal{N}_{i,out} \neq \emptyset$}{
            $\mathcal{N}^{\smax}_{i,out} :=  \{j \in \mathcal{N}_{i,out} : d_j=\max_{\ell \in \mathcal{N}_{i,out}} d_\ell\}$\;
            Node $i$ picks at random a node $j \in \mathcal{N}^{\smax}_{i,out}$\;
            $d^\star=\max\{k,d_j-1,d_i-1\}$\;
            $\mathcal{N}^{\leq}_{ij} :=  \{s \in \mathcal{N}_{j} \setminus \{\mathcal{N}_{i} \cup \{i\}\} : d_s \leq d^\star\}$\;
            \If{$\mathcal{N}_{ij}^{\leq} \neq \emptyset$}{
                $\mathcal{N}^{\smin}_{ij} :=  \{s \in \mathcal{N}^{\leq}_{ij} : d_s=\min_{\ell \in \mathcal{N}^{\leq}_{ij}} d_\ell\}$\;
                Node $i$ picks at random a node $s \in \mathcal{N}_{ij}^{\smin}$\;
                \uIf{$d_j \geq k+1$}{
                    $\E \gets \E \cup \{(i,s)\} \setminus \{(i, j)\}$\;
                }
                \ElseIf{$d_i \leq k$}{
                    Node  $i$ picks at random a value $\beta^{'} \in [0, 1]$\;
                    \If{$\beta' > 1-\beta$}{
                       $\E \gets \E \cup \{(i,s)\}$\;
                    }
                }
            }
        }
        \ElseIf(\textcolor{gray}{\scriptsize \tcp*[f]{Rule 4}}){$r = r_4$}{
            Node $i$ picks at random a node $j\in\mathcal{N}_{i,out}$\;
            \mbox{$\mathcal{N}_{i,out} \gets \mathcal{N}_{i,out} \setminus \{j\}$ and $\mathcal{N}_{j,out} \gets \mathcal{N}_{j,out} \cup \{i\} $\;}
        }
}
\end{DUprotocol}
\endgroup

These operations are executed by first executing Rule~$1$ and then selecting at random one among Rules~{2-3-4}, as illustrated in Fig.~\ref{fig:Algorithm_2}:
\begin{itemize}
    \item \textbf{Rule~1}: node $i$ add edges by repeatedly executing operation~(A1) until its degree becomes $d_i\geq k$;  
    \item \textbf{Rule~2}: node $i$ tries to remove edges by executing operation~(R) if its degree reduces to $d_i< k$;
    \item \textbf{Rule~3}: node $i$ first tries to move an edge by executing operation~(M); then, if it fails, tries (with probability~$1-\beta$) to add an edge by executing operation~(A2).
    \item \textbf{Rule~4}: node $i$ relinquishes ownership of the edge $(i, j)$, where $j \in \mathcal{N}_{i, \text{out}}$, in favor of node $j$.
\end{itemize}
We observe that, unlike the CFOR-REG Protocol, all operations executed by the agents are guaranteed to be non-conflicting. The only scenario worth mentioning is when two non-neighboring nodes $i$ and $j$ attempt to add the same edge $(i,j)$: this situation does not result in a conflict, as the edge can indeed be added (albeit only once), thereby allowing both actions performed by the agents to succeed.
Intuitively, these uncoordinated and conflict-free actions significantly speed up the edge mixing in the network if compared to the DFOR-REG Protocol. On the other hand, the UFA-REG Protocol suffers from possible (even though very unlikely) disconnections and cannot ensure that the graph converges to a random exact regular graph, as shown in the following technical result.
Nevertheless, we provide an empirical validation of the algebraic connectivity of graphs obtained by the UFA-REG Protocol.
\begin{lem}\label{lem:d_min}
    {(Convergence of minimum degree)} By executing the UFA-REG Protocol, the minimum degree $d_{\smin}$ among all agents in the network converges within the interval $[k,\infty)$.
\end{lem}
\begin{pf}
We start by showing that if ${d_{\smin}\leq k}$, then $d_{\smin}$ cannot decrease while the agents execute UFA-REG Protocol.
We analyze all possible operations that may lead to a decrease of one of the agents' degrees, i.e., the removal operation~(R) and the movement operation~(M):
\begin{itemize}
    \item If Rule~2 is selected by an agent $i\in\V$ with degree $d_i\geq k + 1$, the degrees of the nodes can decrease due to the removal operation~(R). In that case, agent $i$ picks at random a neighbor $j\in\mathcal{N}_{i,out}$ with degree $d_j\geq k + 1$, and then removes the edge connecting them.
    Consequently, the degree of agents $i,j$ decreases by $1$, yielding $d_i^+\geq k$ and $d_j^+\geq k$.
   As a result, $d_{\smin}\leq k$ does not decrease.
    \item If Rule~3 is selected by an agent $i\in\V$, the degrees of the nodes can decrease due to the movement operation~(M). In that case, agent $i$ picks at random a neighbor $j\in\mathcal{N}_{i,out}$ with degree $d_j\geq k + 1$ and a $2$-hop neighbor $s\in\mathcal{N}_j$ with degree $d_s\leq \max\{k,d_j-1,d_i-1\}$, and then removes the edge $(i,j)$ and adds the edge $(i,s)$. 
    Consequently, only the degree of agent $j$ decreases, yielding $d_j^+\geq k$.
    Thus $d_{\smin}\leq k$,  and $d_{\smin}$ does not decrease.
\end{itemize}
 
Second, we show that if $d_{\smin}<k$, then it eventually increases at least to $k$. 

If an agent $i\in\V$ such that $d_i<k$ is picked at random, then the selection of Rule~1 ensures that operation~(A1) is repeatedly executed until its degree increases up to $k$.
Indeed, operation~(A1) can add an edge to node $i$ if there is at least a neighbor $j\in\mathcal{N}_i$ which has a neighbor $s\in\mathcal{N}_j$ that is not $i$ itself or a neighbor of $i$, namely
\begin{equation}\label{eq:nij}
\exists j\in\mathcal{N}_{i}:\: \mathcal{N}_{ij} =  \{s \in \mathcal{N}_{j} \setminus (\mathcal{N}_{i} \cup \{i\})\}\neq \emptyset,
\end{equation}
thus increasing its degree by $1$. Agent $i$ repeats this operation whenever Eq. \eqref{eq:nij} holds. If Eq. \eqref{eq:nij} always holds, then, by induction, whenever Rule~1 is selected, the degree of agent $i$ surely increases up to $k$.

We prove by contradiction that Eq. \eqref{eq:nij} always holds.
Assume that ${\mathcal{N}_{ij}=\emptyset}$ for all ${j\in\mathcal{N}_{i}}$, implying that agent $i$ does not have any $2$-hop neighbors.
Consequently, the subgraph containing agent $i$ is a complete graph.
Moreover, if the graph $\G$ is connected then the subgraph containing agent $i$ is $\G$ itself, and thus $\G$ is complete. 
Thus, all agents have the same degree which is equal to $n-1$, yielding the following contradiction
$$
n-1 \overset{(i)}{=} d_{i} \overset{(ii)}{<}k \overset{(iii)}{<} n,
$$
where $(i)$ holds since $\G$ is the complete graph, and $(ii)$ and $(iii)$ hold by assumption. The above inequalities are a contradiction since $k\in\mathbb{N}$ cannot be greater than $n-1$ and also less than $n$.
This proves that the statement in Eq. \eqref{eq:nij} holds if $\G$ is connected. Consequently, agent $i$ can always add an edge to a $2$-hop neighbor, thus increasing its degree $d_i$ until $d_i^+ = k$.
Since all agents are randomly activated with probability $1-\epsilon$,  the minimum degree $d_{\smin}$ eventually becomes such that $d_{\smin}\geq k$. 
\hspace*{\fill}$\qed$
\end{pf}
\begin{lem}\label{lem:d_max}
    {(Convergence of the maximum degree)} By executing the UFA-REG Protocol, if the minimum degree $d_{\smin}$ is within the interval $[k,\infty)$, then the maximum degree $d_{\smax}$ among all agents in the network converges to $[k,k+1]$.
\end{lem}
\begin{pf}
\emph{Based on the result of Lemma~\ref{lem:d_min}, once $d_{\smin}\geq k$ is achieved, we now show that if $d_{\smax}\geq k+1$, then $d_{\smax}$ cannot increase while the agents execute UFA-REG Protocol}.

We analyze all possible operations that may lead to an increase of one of the agents' degrees, i.e., the move operation (M) and the add operation (A2); note that operation (A1) cannot be executed due to $d_{\smin}$.
Both these operations are tried by Rule 3 and are analyzed next:
\begin{itemize}
    \item Operation (M) is executed by agent $i$ if the randomly picked neighbor $j\in\mathcal{N}_{i,out}$ and the randomly picked $2$-hop neighbor $s\in\mathcal{N}_j\setminus\{\mathcal{N}_i\cup \{i\}\}$ are such that $d_s\leq \max\{k,d_j-1,d_i-1\}$. In this case, the edge $(i,j)$ is removed, and the edge $(i,s)$ is added. This means that the degree of node $i$ remains unchanged, $d_i=d_i^+$, the degree of node $j$ decreases, $d_j^+=d_j-1$, and the degree of node $s$ increases,  $d_s^+=d_s+1$. Therefore it holds $d_s^+=d_s+1\leq d_j$, i.e., the maximum degree $d_{\smax}$ does not increase.
    \item Operation (A2) is executed by agent $i$ if $d_i\leq k$ and the randomly picked $2$-hop neighbor $s\in\mathcal{N}_j\setminus\{\mathcal{N}_i\cup \{i\}\}$ satisfies $d_s\leq k$. In this case, the edge $(i,s)$ is added such that the degree of nodes $i,s$ increases, $d_i^+=d_i+1$ and $d_s^+=d_s+1$. Therefore, it holds $d_i^+\leq k+1$ and $d_s^+\leq k+1$, i.e., the maximum degree $d_{\smax}\geq k+1$ does not increase.
\end{itemize}

Second, we prove that if $d_{\smin}\geq k$ and $d_{\smax} > k+1$, then $d_{\smax}$ eventually decreases at least to $k+1$. 

Let $\V_{\smax}$ be the set of nodes with maximum degree $d_{\smax}>k+1$, i.e.,
$
\V_{\smax} = \{i\in\V:d_i=d_{\smax},\: d_i >  k + 1\},
$
and let $\V_{\sint}$ be the set of nodes with degree within ${[k+1,d_{\smax}-1]}$, 
$
{\V_{\sint}=\{\ell\in\V:d_\ell\in[k+1,d_{\smax}-1]\}}.
$
With these notations, if $\V_{\smax}=\emptyset$, then $d_{\smax}\leq k+1$, which is our objective. Thus, we show that there is always a sequence of operations that leads to a decrement in the degree of a node $i^\star \in \V_{\smax}$ such that the set $\V_{\smax}$ eventually gets empty since either one of the following holds:
\begin{enumerate}[label=(\alph*)]
    \item The cardinality of $\V_{\smax}$ decreases as nodes $i^\star$ with the maximum degree $d_{\smax}$ are reduced to $d_{i^\star}^+ = d_{i^\star} - 1 = d_{\smax} - 1$, then $i^\star \notin \V_{\smax}$;
    \item $\V_{\smax}$ contains nodes with decrement degree such that $d_{i^\star}^+=d_{i^\star}-1=d_{\smax}-1>k$ where $i^\star \in \V_{\smax}$, then $d_{\smax} > k$.
\end{enumerate}
In the following discussion, we are not going to consider ownership of the edges since Rule~4 allows a change of ownership of all edges with a non-zero probability.
Assuming that $\V_{\smax}\neq \emptyset$, there are two possible cases:
\begin{enumerate}
    \item \emph{There is a node ${i^\star\in \V_{\smax}}$ whose shortest path to another node in $\V_{\sint}$ has a length greater than or equal to 2}. Node $i^\star$ has only neighbors of degree $k$, otherwise, there would be a shortest path of length 1.
    Let $j^\star$ be the neighbor of $i^\star$ on the shortest path and let $s$ be a neighbor of $i^\star$ that is not on the shortest path and is not a neighbor of $j^\star$, which surely exists due to $d_{i^\star}>d_{j^\star}$. 
    Therefore, node $s$ can execute the move operation (M) selecting node $i^\star$ as neighbor, removing edge $(s,i^\star)$, and selecting  node $j^\star$ as $2$-hop neighbor, adding edge $(s,j^\star)$.
    This leads to a decrease in the degree of node $i^\star\in \V_{\smax}$, i.e., $d_{i^\star}^+=d_{i^\star}-1< d_{\smax}$, and an increase of the degree of node $j^\star$, i.e., $d_{j^\star}^+=d_{j^\star}+1= k+1$. Thus, both $i^\star$ and $j^\star$ enter set $\V_{\sint}$ but not $\V_{\smax}$.
    Thus, either (a) or (b) occur.
    \item \emph{For each $i^\star\in \V_{\smax}$, the shortest path from $i^\star$ to $\V_{\sint}$ has length equal to 1}. Each node $i^\star\in \V_{\smax}$ has at least a neighbor $j^\star\in\V_{\sint}$. 
    Thus, node $i^\star$ can execute the remove operation (R) selecting a neighbor node $j^\star$ and removing edge $(i^\star,j^\star)$.
    This leads to a decrease in the degree of both nodes $i^\star$ and $j^\star$.
    Thus, either (a) or (b) occurs.
\end{enumerate}
By repeating the operations above described, the set $\V_{\smax}$ will eventually become empty, i.e., $d_{\smax}\leq k+1$. This completes the proof of Lemma \ref{lem:d_max}.
\hspace*{\fill}$\qed$
\end{pf}
\begin{thm}\label{thm:ufa-reg}
Consider an RGP due to the execution of the UFA-REG Protocol starting from a connected graph $\G(0)$ with $n\geq k$ nodes and even $k\in\{4,6,8,\ldots\}$. Then, it almost surely exists a $t^\star \in \nat$ such that the graph $\G(t)$ is approximate $k\text{-regular}$ for all times $t\geq t^\star$, i.e., 
$$ 
\pr(\exists t^\star>0, \forall t\geq t^\star: \G(t) \text{ is app. $k$-regular}) = 1.
$$
\end{thm}
\begin{pf}
    We study the behavior of the minimum $d_{\smin}$ and maximum $d_{\smax}$ degrees among all the agents in the network, showing first that $d_{\smin}$ eventually increases and remains greater than or equal to $k$, and, secondly, that $d_{\smax}$ eventually decreases and remains lesser than or equal to $k+1$, thus making $\G'$ an approximate $k$-regular graph.   
    Lemma \ref{lem:d_min} and \ref{lem:d_max} imply that all the degrees converge to $[k,k+1]$, with a finite number of operations on the graph, completing the proof. 
\hspace*{\fill}$\qed$
\end{pf}

\begin{figure}[!b]
    \centering    \includegraphics[width=0.43\textwidth]{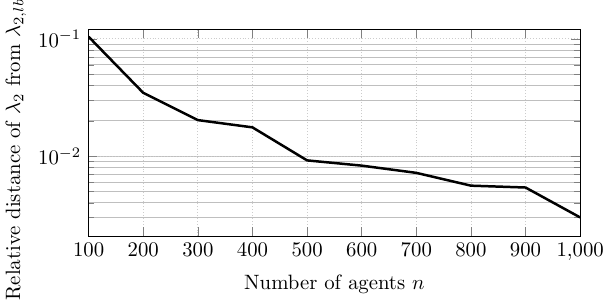}
    \caption{Plot of $|\lambda_2 -\lambda_{2,lb}|/\lambda_{2,lb}$ for graphs generated by the UFA-REG Protocol with  $\lambda_{2,lb}$ as in Proposition~\ref{pro:specgap} with $k=\lfloor \sqrt{n}\rfloor$.}
    \label{fig:perc_err}
\end{figure}

We now empirically validate that the algebraic connectivity of a graph generated by the UFA-REG Protocol is greater in expectation than the lower bound for a random $k\text{-regular}$ graph as characterized in Proposition~\ref{pro:specgap}.
We run experiments on networks with an increasing number of agents ${n\in\{100,200,\ldots,1000\}}$ and select a degree of regularity $k=\lfloor \sqrt{n}\rfloor$ proportional to the number of nodes.
Figure~\ref{fig:perc_err} shows the relative distance of the algebraic connectivity $\lambda_2$ from the lower bound $\lambda_{2,lb}$, i.e., $(\lambda_2-\lambda_{2,lb})/\lambda_{2,lb}$, averaged over 1000 different instances of the problem.
The results show that the lower bound for a connected random $k\text{-regular}$ graph provided by Proposition~\ref{pro:specgap} is always achieved (thus showing it actually preserves connectivity) and the relative distance to it decreases with the size of the network.

\begin{figure*}[!t]
\centering
\includegraphics[width=0.235\textwidth]{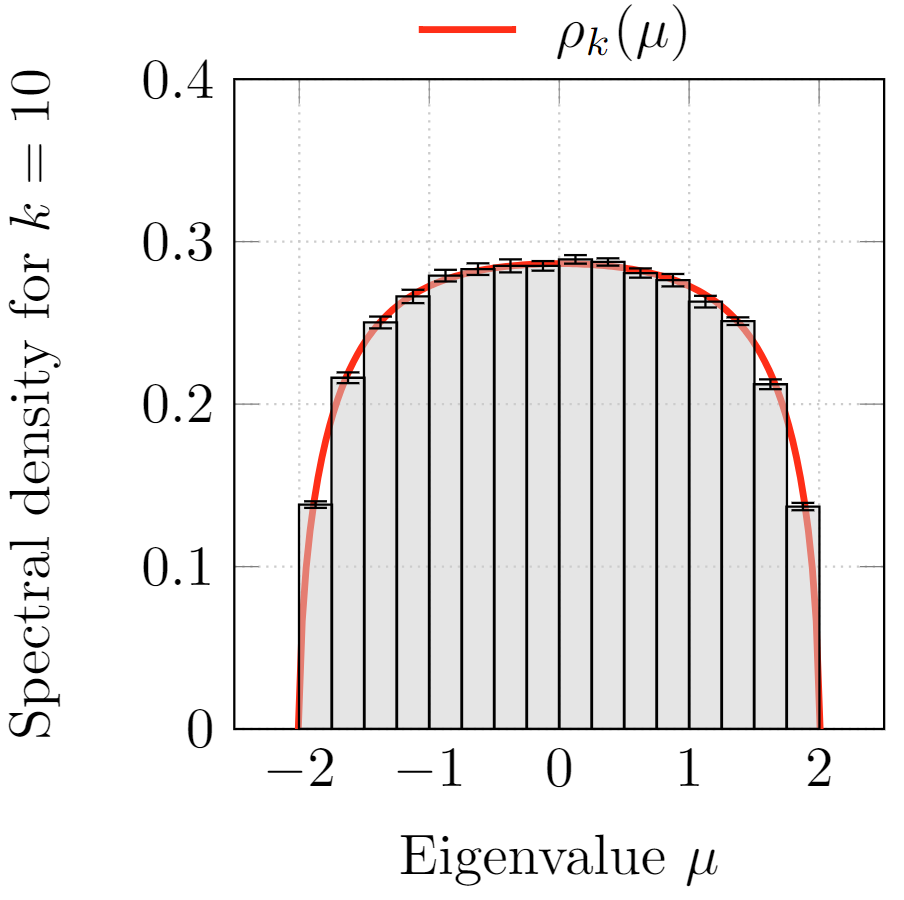}
\includegraphics[width=0.235\textwidth]{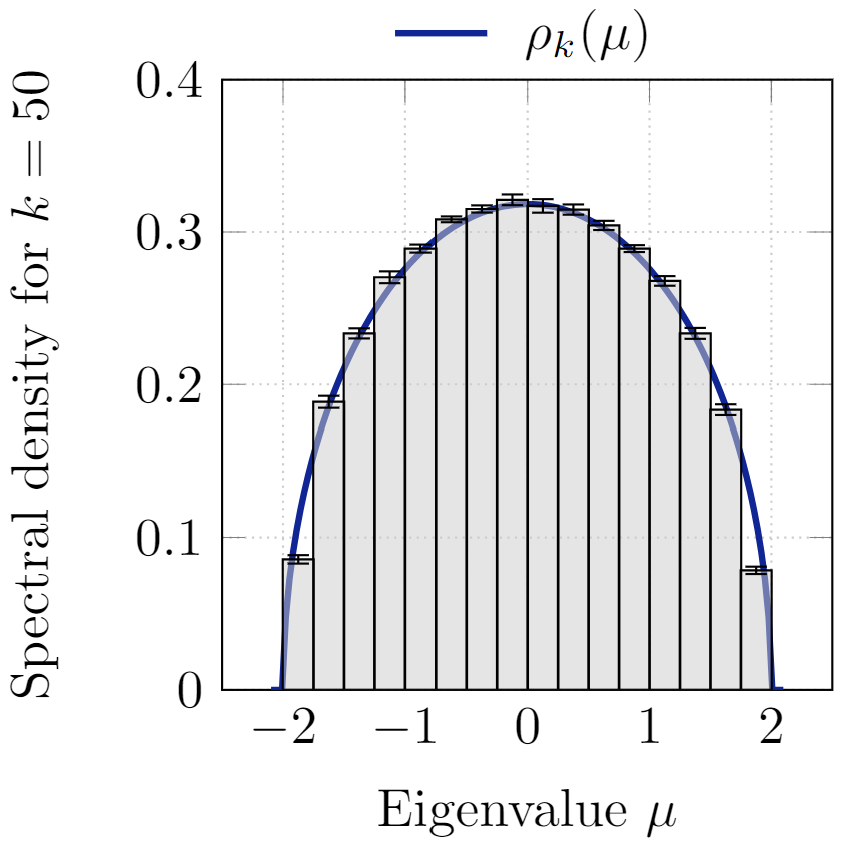}
\includegraphics[width=0.235\textwidth]{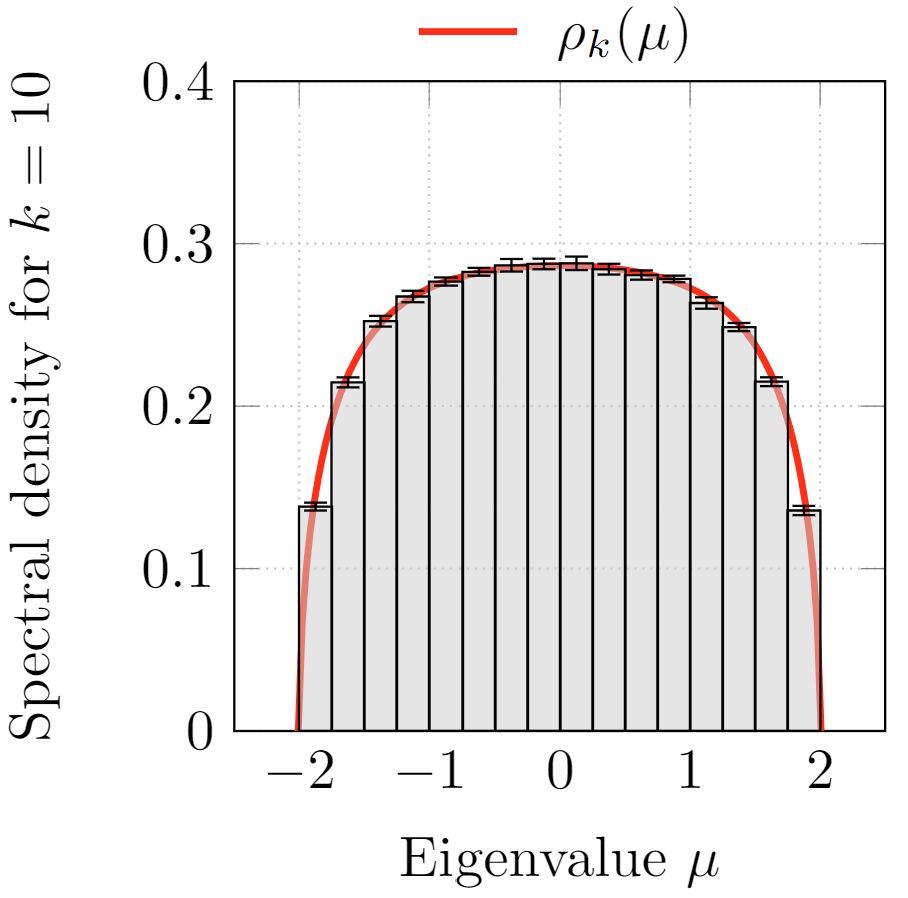}
\includegraphics[width=0.235\textwidth]{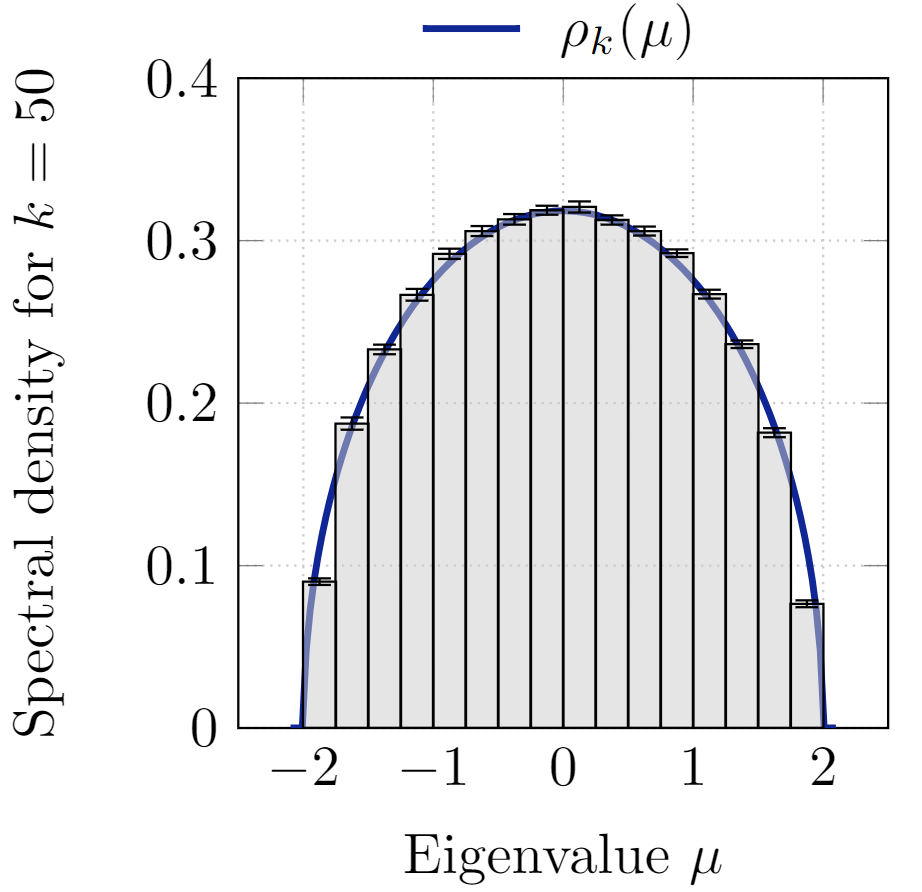}
\caption{Empirical eigenvalue density histogram of the normalized adjacency matrix of random (approximate) regular graphs with $n=1000$ nodes generated by (left) the CFOR-REG Protocol and (right) the UFA-REG Protocol:
the red curve represents the density $\rho_k$ expected for large regular graphs $n\rightarrow\infty$ (see Proposition~\ref{pro:dfix});
the blue curve represents the density $\rho_{sc}$ expected for large regular graphs and high values of the regularity degree $n,k\rightarrow \infty$ (see Proposition~\ref{pro:dinc}).}
\label{fig:dfix}
\end{figure*}

\subsection{Validation of the empirical spectral distribution}\label{sec:specdisttheo}
We now validate the \textit{empirical spectral distribution} (ESD) of graphs generated with the proposed protocols with that of random regular graphs. 
The eigenvalues $\mu_i$ of the adjacency matrix $A_n$ associated with a random graph $\G$ are samples of independent, identically distributed random variables.
In these terms, the ESD $P_{A_n}:\rea\rightarrow [0,1]$ of the matrix $A_n$ is an estimate of the \textit{cumulative distribution function} $P(x):\rea\rightarrow [0,1]$ that generates its eigenvalues,
$
{P_{A_n}(x) = \frac{1}{n}\abs{\{i:\mu_i\leq x\}}},
$
where $\abs{\cdot}$ denotes the cardinality of a set.
In simple terms, the distribution $P(x)$ is the probability that an eigenvalue takes a value less than or equal to $x$, and the ESD $P_{A_n}(x)$ is an approximation of this probability given the realization~$A_n$.
Moreover, by the strong law of large numbers, the ESD $P_{A_n}(x)$ almost surely converges to $P(x)$ for $n\rightarrow \infty$.
Another important concept is the relative likelihood that an eigenvalue is equal to a specific value, which is given by the \textit{probability density function} $\rho:\rea \rightarrow \rea$, defined by
$$
\lim_{n\rightarrow \infty} P_{A_n}(x) = P(x) = \int_{-\infty}^x \rho(x) dx.
$$
A first important characterization of the ESD of random ${k\text{-regular}}$ graph has been provided by McKay in~\cite{b27}, building upon~\cite{b28}, by considering the case of a fixed degree of regularity $k$ in the limit of $n\rightarrow \infty$.
\begin{prop}\label{pro:dfix}
Let $A_n$ be the adjacency matrix of a random ${k\text{-regular}}$ connected graph with $n$ nodes. 
When ${n}$ approaches to ${\infty}$, the ESD of the normalized adjacency matrix
$
A_{\sigma} = A_n/\sigma$ with $\sigma = \sqrt{k-1}
$
approaches the distribution with density
$$
\rho_{k}(x) = \begin{cases}
\frac{k^2-k}{2\pi(k^2-kx^2+x^2)}\sqrt{4-x^2} & \text{\rm{if }} \abs{x}\leq 2,\\0 & \text{otherwise}.
\end{cases}
$$
\end{prop}
It can be verified that when $k\rightarrow\infty$, the ESD of $A_\sigma$ in Proposition~\ref{pro:dfix} converges to a distribution with semicircle density. 
Following this idea, Tran, Vu, and Wang in~\cite[Theorem 1.5]{b29} proved the following result.
\begin{prop}\label{pro:dinc}
Let $A_n$ be the adjacency matrix of a random ${k\text{-regular}}$ graph with $n$ nodes. 
When $k,n\rightarrow\infty$, the ESD of the normalized adjacency matrix
$
A_{\sigma} = A_n/\sigma$ with $\sigma = \sqrt{k-k^2/n}
$
approaches the distribution with semicircle density
$$
\rho_{sc}(x) = \begin{cases}
\frac{1}{2\pi}\sqrt{4-x^2}, & \text{\rm{if }} \abs{x}\leq 2,\\
0, & \text{otherwise}.
\end{cases}
$$
\end{prop}
We note that the normalization in Proposition~\ref{pro:dfix} is a special case of that of Proposition~\ref{pro:dinc} for $n=k^2$.
The semicircle density in Proposition~\ref{pro:dinc} is mostly known due to the \textit{Wigner's semicircle law}~\cite{b30}, which describes the limiting spectral distribution of large random matrices with independent, identically distributed entries. 

We run Monte-Carlo simulations in large networks of ${n=1000}$ nodes considering two different scenarios:
\begin{itemize}
    \item A small degree of regularity $k=10$, for which the spectral distribution is given in Proposition~\ref{pro:dfix};
    \item A high degree of regularity $k=50$, for which the spectral distribution is given in Proposition~\ref{pro:dinc}.
\end{itemize}
The comparison is carried out by computing the eigenvalues of the normalized adjacency matrix $A_{\sigma} = A_n/\sigma$ (where $\sigma$ is a proper normalization factor) and comparing their empirical density distribution with the distributions characterized in Propositions~\ref{pro:dfix}--\ref{pro:dinc} for $k$-regular graphs.
Figure~\ref{fig:dfix} shows the results of the above two cases, averaged over 10 different instances for each problem.
It can be noticed that in both cases, and for both protocols, the limiting distribution approaches the distribution of a random $k\text{-regular}$ graph according to Propositions~\ref{pro:dfix}--\ref{pro:dinc}. 
For the sake of completeness, we detail how the empirical spectral densities histograms in Figure~\ref{fig:dfix} are obtained: 1) Divide the range of values $[-2,2]$ into consecutive, non-overlapping intervals: we selected $16$ intervals of fixed width equal to $0.25$; 2) Compute the eigenvalues of the normalized adjacency matrix and count how many of them fall into each interval; 3) Plot a bar for each interval with a height equal to the corresponding number of eigenvalues divided by their total number.

\section{Numerical simulations}
\label{sec:numerical}
We compare the performance of the proposed protocols with other two algorithms in the state-of-the-art:
\begin{itemize}
    \item Algorithm 1 in \cite{Zohreh22}: this algorithm allows to choose arbitrarily the degree of regularity $k$;
    \item Algorithm 2 in \cite{Yazicioglu2015formation}: this algorithm does not allow to choose the degree of regularity, which is constrained within an interval defined by the initial average degree, namely $k\in[d_{\savg}(\G),d_{\savg}(\G)+2]$.
\end{itemize}

\subsection{Numerical comparison of performance}
To make a fare comparison with Algorithm~2 in \cite{Yazicioglu2015formation}, we consider a network $\G=(\V,\E)$ with $n=\abs{\V}=1000$ agents whose communication graph has an initial average degree equal to $d_{\savg}(\G)=49.2$. Then, we set the desired degree of regularity equal to $k=50$, an high nodes' activation probability $1-\varepsilon=0.99$, and a low probability $\beta=0.01$ of adding edges when not strictly needed.
In this way, the expected graphs generated by the four algorithms are:
\begin{itemize}
    \item CFOR-REG Protocol produces a random $k\text{-regular}$ graph with $k=50$;
    \item UFA-REG Protocol produces an approximate $k\text{-regular}$ graph with $k=50$, which well approximate a random regular graph;
    \item Algorithm 1 in \cite{Zohreh22} produces a random approximate $k\text{-regular}$ graph with $k=50$;
    \item Algorithm 2 in \cite{Yazicioglu2015formation} produces a random $k\text{-regular}$ graph with $k\in\{50,51\}$.
\end{itemize}
%
%
\begin{figure}[!t]
    \centering
    \includegraphics[width=0.48\textwidth]{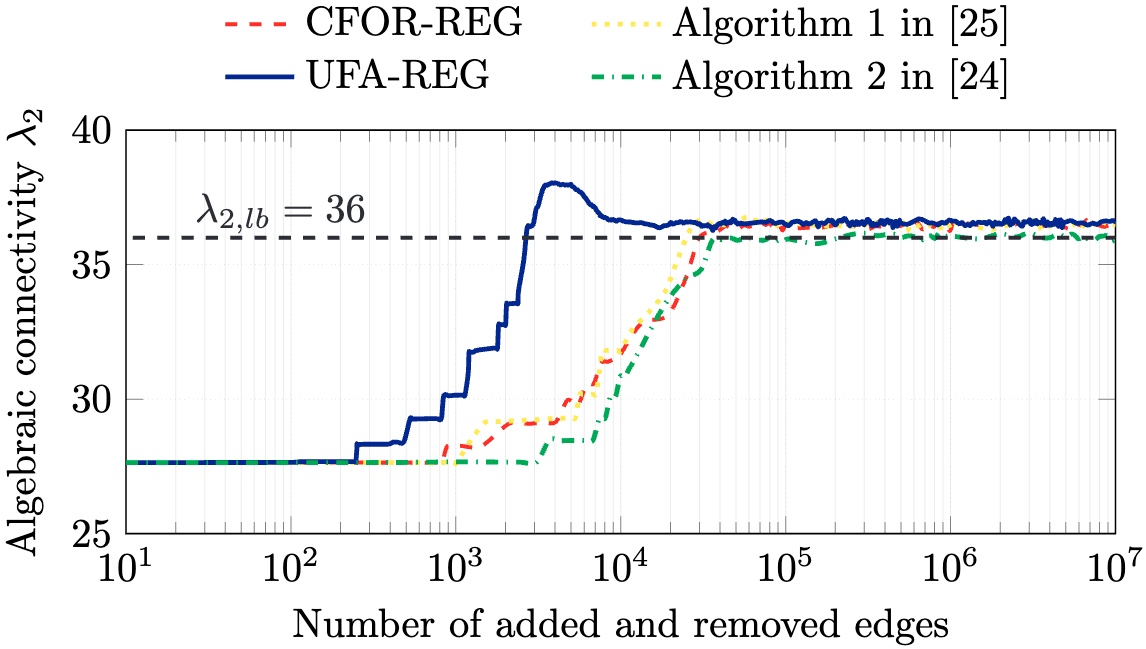}
    \includegraphics[width=0.48\textwidth]{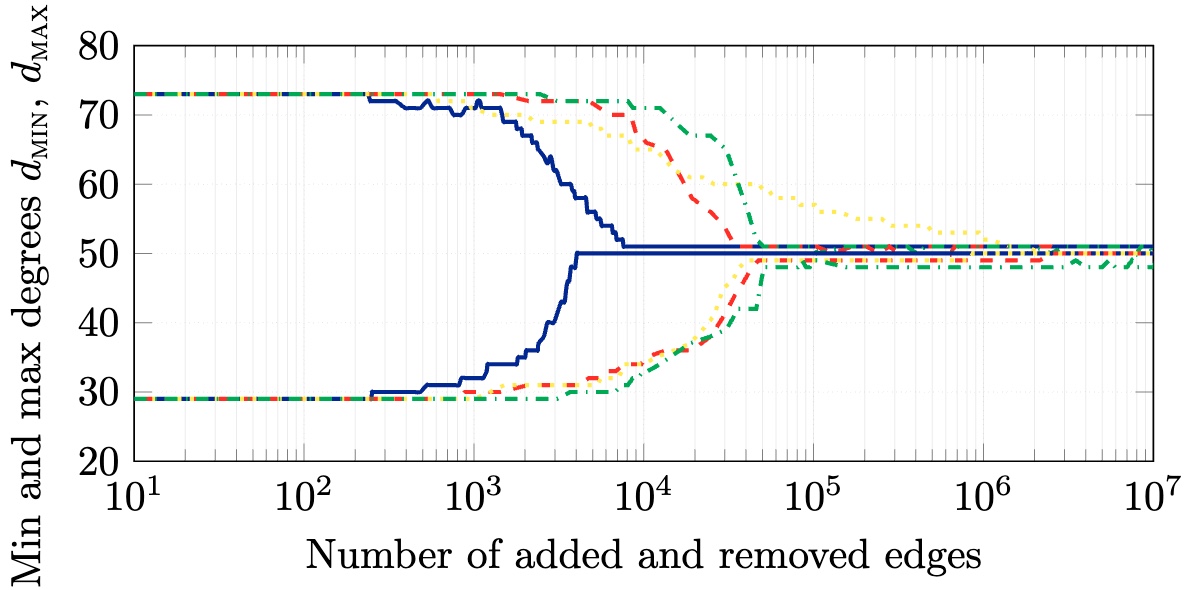}
    \caption{Comparison with the state-of-the-art of the evolution of (top) the algebraic connectivity and (bottom) the maximum/minimum degrees, against the number of added and removed edges, during the execution of the algorithms.}
    \label{fig:comparison_1000}
\end{figure}

Figure~\ref{fig:comparison_1000} shows the evolution of the algebraic connectivity $\lambda_{2}$ (on the top) and of the minimum $d_{\smin}$ and maximum $d_{\smax}$ degrees (on the bottom) of the network while executing the proposed protocols and those proposed in the state-of-the-art.
In order to make the comparison fair, we plot these evolutions against the number of added and removed edges during the execution of the algorithms, which we call \virg{\textit{actions}}: adding or removing an edge counts as one action, moving an edge counts as two actions, and exchanging two neighbors counts as four actions.

The numerical simulations reveal that the graphs generated by all algorithms increase their algebraic connectivity $\lambda_2$, approaching and then exceeding the lower bound $\lambda_{2,lb}:=k-2\sqrt{k-1} = 36$ for random $k\text{-regular graphs}$ for $k=50$, see Proposition~\ref{pro:specgap}.
It can be noticed that the UFA-REG Protocol is the fastest one, allowing to achieve the desired algebraic connectivity by adding/removing less than $3\cdot 10^3$ edges, while the other algorithms about $3\sim5\cdot 10^4$ actions, which is one extra order of magnitude.
Moreover, the proposed protocols are also the fastest ones in achieving an (approximate) $k$-regular graph, requiring less than $5\cdot 10^4$ actions, while Algorithm 1 in \cite{Zohreh22} needs about $10^6$ actions and Algorithm 2 in \cite{Yazicioglu2015formation} did not converge to a regular graph even with $10^7$ actions.

\subsection{Resilience against attacks}
We test the proposed UFA-REG Protocol and the CFOR-REG Protocol when the network is under attack.
We consider an attacker that has the objective of disconnecting the network by carrying out DoS attacks at the nodes of the network, i.e., causing their failure and disconnection from the network. 
The amount of time needed by the attacker to select a node in the network and then perform the attack is strictly greater than zero, such that there is a minimum number of iterations $\Upsilon>0$ that the algorithms can perform before the attack succeeds. In our simulations, we assume that the attacker is sufficiently fast that it can perform an attack every $\Upsilon=10$ iterations of the algorithms.
Moreover, we assume that the attacker has full knowledge of the network, enabling the identification of nodes that, when removed, significantly reduce the connectivity and resilience to disconnections of the network.
To simulate this selection process, the attacker exploits the eigenvector of the Laplacian matrix associated with the algebraic connectivity, known as the Fiedler eigenvector. The sign of each entry of the Fiedler eigenvector denotes which side of a partition the corresponding node belongs to when the graph is split into two parts based on the Fiedler eigenvector. 
Thus, the attacker selects one of the nodes that have the highest number of neighbors with opposite Fiedler eigenvector sign entry, i.e., in the opposite partition. These nodes are likely to be close to the \virg{center} of the graph and their removal may potentially bisect the graph.

\begin{figure}[!t]
    \centering
    \includegraphics[width=0.48\textwidth]{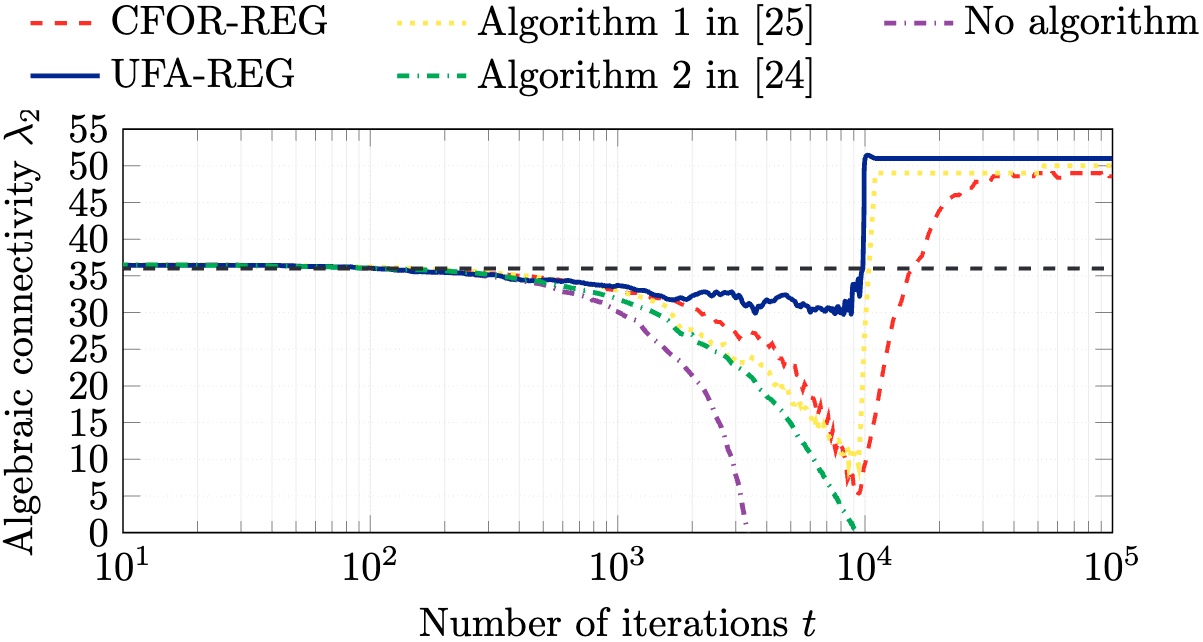}
    \caption{Evolution of $\lambda_2$ in initially $50$-regular graphs of $n=1000$ nodes from which one node is disconnected every $\Upsilon=10$ steps.}
    \label{fig:opennet}
\end{figure}

Figure~\ref{fig:opennet} compares the results of the attack on a random $50$-regular network of $n=1000$ agents that employ one of the proposed protocols, or one of the state-of-the-art algorithms in \cite{Zohreh22,Yazicioglu2015formation}, with the case in which no algorithm for self-organizing the network is employed. 
The results reveal that the UFA-REG Protocol outperforms all others, allowing a steadily increasing algebraic connectivity of the overall network and, in turn, avoiding disconnections.
The only algorithm that could not avoid the network disconnection is Algorithm 2 in \cite{Yazicioglu2015formation}; on the other hand, it was able to delay the disconnection until more than $90\%$ of the nodes have been attacked.
Finally, the CFOR-REG Protocol and Algorithm 1 in \cite{Zohreh22} are able to keep the network connected, even though the algebraic connectivity decreased during the attack.
When no self-organization logic is implemented, the attack successfully disconnects the network by attacking less than $40\%$ of the nodes.

\begin{rem}
Note that the attack is stopped when the network reached a number of $k+1$ nodes, for which a $k\text{-regular}$ graph becomes the complete graph without self-loops, whose algebraic connectivity is $\lambda_2= n+1=50$.
\end{rem}


\section{Conclusions}
\label{sec:conclu}

This manuscript presents two distributed protocols that enable the self-organization of any network topology into one with desired algebraic connectivity and bounded degrees, despite node failures due attacks.
The strategy employed by the protocols is that of steering the graph topology toward that of a random (approximate) $k\text{-regular}$ graph, where $k$ is a free design parameter known by each agent representing the desired node degree.
For the CFOR-REG Protocol, a rigorous formal analysis of its convergence properties is presented while for the UFA-REG Protocol, which is characterized by improved performance, a numerical validation through Monte Carlo simulations is provided.
Future work will focus on developing methods to self-organize graphs into  $r$-robust graphs for networks of multi-agent systems and peer-to-peer networks.

\printbibliography

\end{document}